\documentclass[11pts]{aa} 

\usepackage[utf8]{inputenc} 

\usepackage{geometry} 
\usepackage{graphicx} 
\usepackage{hhline}

\usepackage{booktabs} 
\usepackage{array} 
\usepackage{paralist} 
\usepackage{verbatim} 
\usepackage{subfig} 
\usepackage{natbib}
\usepackage{threeparttable}

\usepackage{fancyhdr} 
\def\farcs{\hbox{$.\!\!^{\prime\prime}$}}  

\usepackage{sectsty}
\allsectionsfont{\sffamily\mdseries\upshape} 

\usepackage[nottoc,notlof,notlot]{tocbibind} 
\usepackage[titles,subfigure]{tocloft} 

\usepackage{color}
\usepackage{sidecap}

\usepackage{multirow}
\usepackage{tabulary}
\usepackage{tabularx}

\usepackage{booktabs}
\usepackage{lipsum}
\usepackage{afterpage}

\newcolumntype{P}[1]{>{\centering\arraybackslash}p{#1}}

\usepackage{natbib,twoopt}
\usepackage[breaklinks=true]{hyperref} 
\bibpunct{(}{)}{;}{a}{}{,}             
\makeatletter
  \newcommandtwoopt{\citeads}[3][][]{\href{http://adsabs.harvard.edu/abs/#3}%
    {\def\hyper@linkstart##1##2{}%
     \let\hyper@linkend\@empty\citealp[#1][#2]{#3}}}
  \newcommandtwoopt{\citepads}[3][][]{\href{http://adsabs.harvard.edu/abs/#3}%
    {\def\hyper@linkstart##1##2{}%
     \let\hyper@linkend\@empty\citep[#1][#2]{#3}}}
  \newcommandtwoopt{\citetads}[3][][]{\href{http://adsabs.harvard.edu/abs/#3}%
    {\def\hyper@linkstart##1##2{}%
     \let\hyper@linkend\@empty\citet[#1][#2]{#3}}}
  \newcommandtwoopt{\citeyearads}[3][][]%
    {\href{http://adsabs.harvard.edu/abs/#3}
    {\def\hyper@linkstart##1##2{}%
     \let\hyper@linkend\@empty\citeyear[#1][#2]{#3}}}
\makeatother

\title{Microphysics and dynamics of the Gamma-Ray Burst 121024A}

\author{K. Varela$^{1}$ \and H. van Eerten$^{1}$\thanks{Alexander von Humbolt Fellow} \and  J. Greiner$^{1,2}$ \and P. Schady$^{1}$ \and J. Elliott$^{1,3}$ \and V. Sudilovsky$^{1,3}$ \and T. Kr\"{u}hler$^{1,4}$ \and A.J. van der Horst$^{5}$ \and J. Bolmer$^{1,6}$ \and F. Knust$^{1}$ \and C. Agurto$^{4}$ \and F.  Azagra$^{4}$ \and A. Belloche$^{7}$ \and F. Bertoldi$^{8}$ \and C. De Breuck$^{9}$ \and C. Delvaux$^{1}$ \and R. Filgas$^{10}$ \and J. F. Graham$^{1}$ \and D. A. Kann$^{11}$ \and S. Klose$^{11}$  \and K. M. Menten$^{7}$ \and A. Nicuesa Guelbenzu$^{11}$ \and A. Rau$^{1}$ \and A. Rossi$^{11,12}$ \and S. Schmidl$^{11}$ \and F. Schuller$^{7}$ \and T. Schweyer$^{1,6}$ \and M .Tanga$^{1}$ \and A. Weiss$^{7}$ \and P. Wiseman$^{1}$ \and F. Wyrowski$^{7}$}

\institute{Max-Planck-Institut f\"{u}r Extraterrestrische Physik, Giessenbachstra{\ss}e, 85748, Garching, Germany \\
e-mail: kvarela@mpe.mpg.de
\and Excellence Cluster Universe, Technische Universit\"{a}t M\"{u}nchen,  Boltzmannstra{\ss}e 2, 85748, Garching, Germany
\and Astrophysics Data System, Harvard-Smithonian Center for Astrophysics, Garden St. 60, Cambridge, MA 02138, U.S.A.
\and European Southern Observatory, Alonso de C\'ordoba 3107, Vitacura, Casilla 19001 Santiago 19, Chile
\and Department of Physics, The George Washington University, 725 21st Street NW, Washington, DC 20052, USA
\and Technische Universit\"{a}t M\"{u}nchen, Physik Dept., James-Franck-Str., 85748 Garching, Germany
\and Max-Planck-Institut f\"{u}r Radioastronomie, Auf dem H\"{u}gel 69, 53121 Bonn, Germany
\and Argelander-Institut f\"{u}r Astronomie, Auf dem H\"{u}gel 71, 53121 Bonn, Germany
\and European Southern Observatory, Schwarzschild-Str. 2, 85748 Garching, Germany
\and Institute  of  Experimental  and  Applied  Physics, Czech  Technical University in Prague, Horska 3a/22, 128 00 Prague 2, Czech Republic
\and Th\"{u}ringer Landessternwarte Tautenburg, Sternwarte 5, 07778 Tautenburg, Germany
\and INAF-IASF Bologna, Area della Ricerca CNR, via Gobetti 101, 40129 Bologna, Italy
}

\date{} 
\usepackage{amstext}

\begin{document}

\abstract {}{The aim of the study is to constrain the physics of gamma-ray bursts (GRBs) by analysing the multi-wavelength afterglow data set of GRB 121024A that covers the full range from radio to X-rays.}{Using multi-epoch broad-band observations of the GRB 121024A afterglow, we measured the three characteristic break frequencies of the synchrotron spectrum. We used six epochs of combined XRT and GROND data to constrain the temporal slopes, the dust extinction, the X-ray absorption, and the spectral slope with high accuracy. Two more epochs of combined data from XRT, GROND, APEX, CARMA, and EVLA were used to set constraints on the break frequencies and therefore on the micro-physical and dynamical parameters.}{The XRT and GROND light curves show a simultaneous and achromatic break at around $49$ ks. As a result, the crossing of the synchrotron cooling break is no suitable explanation for the break in the light curve. The multi-wavelength data allow us to test two plausible scenarios explaining the break:  a jet break, and the end of energy injection. The jet-break scenario requires a hard electron spectrum, a very low cooling break frequency, and a non-spreading jet. The energy injection avoids these problems, but requires $\epsilon_e > 1$ ($k=2$), spherical outflow, and $\epsilon_B < 10^{-9}$.}{In light of the extreme microphysical parameters required by the energy-injection model, we favour a jet-break scenario where $\nu_m < \nu_{sa}$  to explain the observations. This scenario gives physically meaningful microphysical parameters, and it also naturally explains the reported detection of linear and circular polarisation.}


\maketitle 


\section{Introduction}

Gamma-ray bursts (GRBs) are the most luminous phenomena detected so far in the Universe \citep{1973ApJ...182L..85K}. They consist of pulses of gamma-rays emitted in a short time interval (milliseconds to hours) with an isotropic equivalent energy release of $10^{50}-10^{54}$ erg. This is followed by fading multi-wavelength emission (from X-rays to radio), known as a GRB afterglow \citep{1997Natur.387..783C,1997Natur.386..686V}. In the standard afterglow model, the dominant process during the afterglow phase is synchrotron emission from shock-accelerated electrons in a collimated relativistic blast wave interacting with the external medium \citep{1997meszarosapj}. This unique featureless intrinsic spectrum makes GRB afterglows perfect events to study physical processes under extreme conditions (e.g. Fermi acceleration). \\

The observed synchrotron spectrum is composed of four power-law segments joined at three main break frequencies. Each break frequency yields specific and correlated constraints on the acceleration processes in the shock region and the dynamics and geometry of the relativistic outflow. Assuming an initially self-similar evolution for the relativistic blast wave in the afterglow stage \citep{BM1976}, the properties of the synchrotron spectrum can be expressed in terms of constraints on the model parameters \citep{1997MNRAS288wijers,1998ApJ...497L..17S}. The dynamics of the outflow are dictated by the isotropic equivalent energy in the afterglow phase $E_{\rm{iso}}$, the circumburst medium density $n = Ar^{-k}$ (where A is a scale factor, $r$ is the radial distance from the source, and k is the power-law slope with $k = 0$ or $k = 2$ for inter-stellar medium (ISM) or stellar wind-like medium-density profiles, respectively) and, when the jet nature of the outflow becomes apparent, the jet half-opening angle $\theta_0$ \citep{1999ApJ...525..737R}. The micro-physics of this synchrotron emission can be captured in a simplified manner using the post-shock energy fraction in accelerated electrons $\epsilon_{\rm{e}}$, the energy fraction in the magnetic field $\epsilon_{\rm{B}}$ , and the power-law index $p$ of the non-thermal electron population.\\


A snapshot of the spectral energy distribution (SED) covering all break frequencies gives a number of observational constraints equal to the number of model parameters. This scenario presents an ideal case in terms of model simplicity and data availability.  In practice, the sample of bursts for which all spectral breaks could be simultaneously determined is still small ($\text{about
five,}$  e.g. \citealt{Panaitescu2002,2003ApJ...597..459Y,2005AA...440..477R,2008ApJ...683..924C,Cenko2010}). This indicates the importance, in terms of sample statistics, of increasing the sample of bursts with simultaneous afterglow detections across the broadband spectrum. Common solutions when not all spectral breaks can be probed simultaneously are to extrapolate light curves in time and/or to fix one or more of the model parameters. Motivated by an apparent correlation \citep{2001ApJ...562L..55F,Panaitescu2002}, $E_{iso}$ has been equated to the total isotropic equivalent energy release in gamma rays $E_{\gamma,iso}$ (e.g. \citealt{1999ApJ...519L.155D,2003BASI...31...19P}). The magnetic field energy density, $\epsilon_{B}$ has been fixed to a standard value (e.g. \citealt{2000ApJ...537..191F,Cenko2010}), or linked to the shock-accelerated electron energy, $\epsilon_e$ (e.g. \citealt{2006ApJ...651L...9M,2009MNRAS.394.2164V}). A standard value for the circumburst density has also been assumed (e.g. \citealt{2004ApJ...606..369C}). In these cases, the implications of the derived model parameters are conditional on the additional assumption(s).\\

Here, we present the analysis of the simultaneous multi-wavelength observations of GRB 121024A. It was detected with the \textit{Swift} satellite and had a redshift $z$ = 2.30 measured with the X-shooter spectrograph at the Very Large Telescope (VLT) \citep{2012GCN..13890...1T}. It was followed up by different instruments in the radio to the X-ray regimes over several days. Linear and circular optical polarisation observations of the afterglow were taken \citep{2014Natur.509..201W}. We report on the analysis of the broad-band SED of this afterglow, including X-ray, optical/NIR, sub-mm, and radio data. From these simultaneous broad-band observations, we derive constraints on the micro-physical and dynamical parameters of the GRB afterglow. \\

We provide a brief summary of the observations and relevant details of the data reduction in Sect. \ref{observations}. We then describe in Sect. \ref{analysis} a model-independent analysis of the data, starting with the description of the X-ray and optical/NIR light curves, followed by the description of the SED including effects of dust extinction and gas absorption at these wavelengths. In Sect. \ref{theory} we include radio and sub-mm data to study the broad-band SED in the framework of jet break and energy-injection scenarios. We derive all the micro-physical and dynamical parameters based on the standard afterglow model assumptions. In Sect. \ref{dis} we discuss our results and compare the different viable scenarios. Finally, we conclude and summarise our results in Sect. \ref{sum}.

\section{Observations and data reduction}
\label{observations}

\subsection{\textit{Swift}}
On 2012 October 24 at $T_0$ = 02:56:12 UT, the \textit{Swift} Burst Alert Telescope \citep[BAT,][]{2005SSRv..120..143B} triggered and located GRB 121024A \citep{2012GCNpagani}. \textit{Swift} slewed immediately to the burst, and the observations with the X-Ray Telescope \citep[XRT,][]{2005SSRv..120..165B} started 93 sec after the trigger. The observations were made in windowed timing (WT) mode during the first $242$ s and then were carried out in photon counting (PC) mode \citep{2012GCN..13892...1P}. The initial flux in the $0.2-10$ keV band was $1.1\times10^{-9}$ erg cm$^{-2}$ s$^{-1}$. The \textit{Swift}/XRT light curve and spectral data were obtained from the XRT repository \citep{2007AA469Evans,2009MNRAS.397.1177Evans}. The afterglow was located RA, Dec (J2000) = 04:41:53.28, -12:17:26.8 with an uncertainty of 0\farcs8 \citep{2012GCNpagani} by the \textit{Swift}/UVOT, with a magnitude in the $b$ band of $18.4\pm0.2$ \citep{2012GCN..13901...1H}.  

\subsection{GROND}
\begin{figure}[!ht]
\centering
\includegraphics[width=0.5\textwidth]{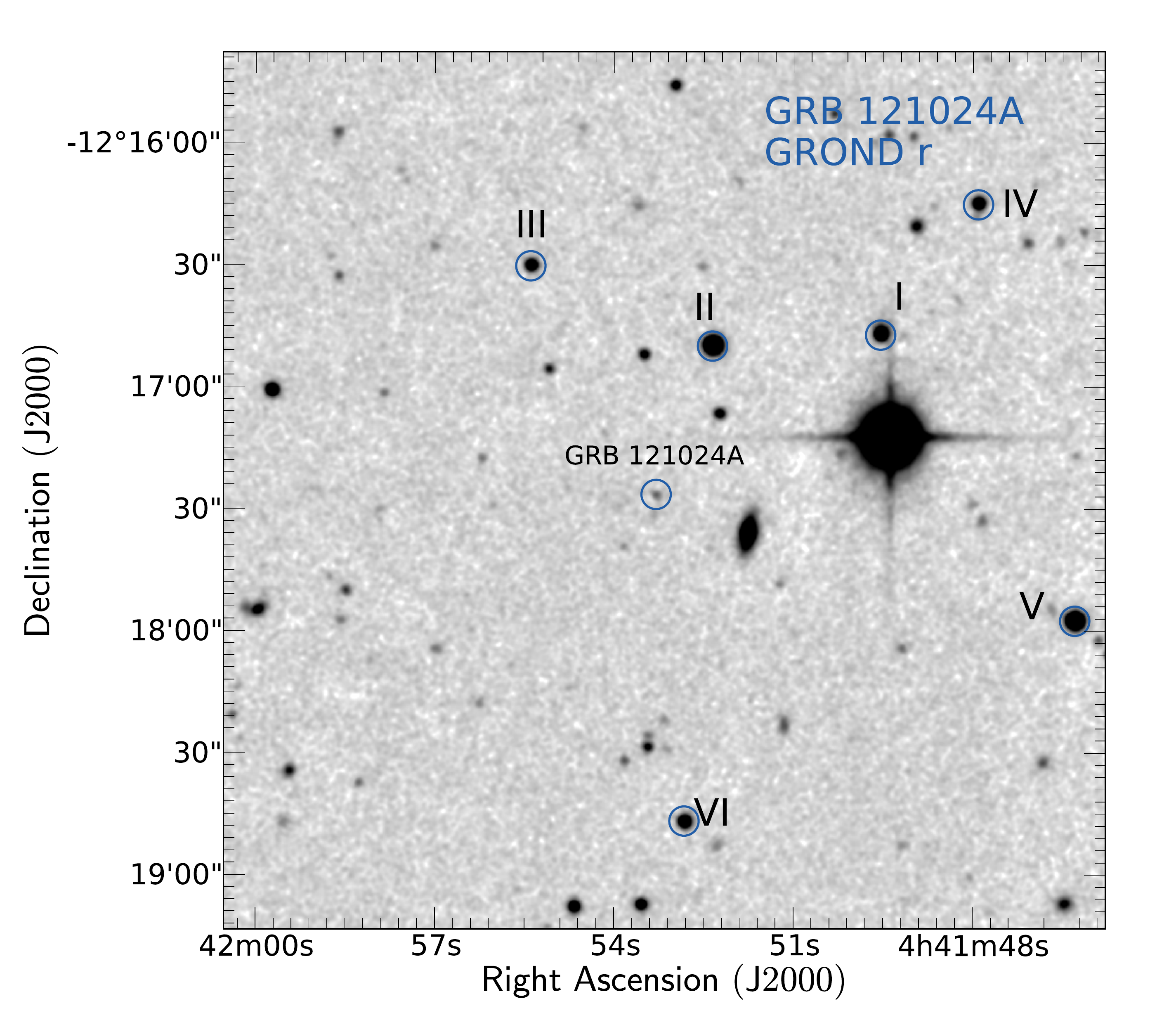}

\caption{GROND $r'$-band finding chart. The secondary stars used for the calibration are labelled I-VI and are reported in Table \ref{refstars1}. North is up and east to the left.}
\label{FCgrb121024A}
\end{figure}

The Gamma-Ray burst Optical Near-infrared Detector - GROND \citep{Greiner2008} mounted at the Max-Planck-Gesellschaft (MPG) 2.2m telescope located at ESO La Silla observatory, Chile, was designed as a GRB follow-up instrument. It provides simultaneous data in seven bands in a wavelength range from 400-2400 nm ($g'r'i'z'JHK_{\rm{s}}$). GROND observations started 2.96 hours after the \textit{Swift} trigger \citep{2012GCNknust} and continued for the next 3.8 hours during the first night. The afterglow was detected in all seven bands at the position RA, Dec (J2000) = 04:41:53.30, -12:17:26.5 with an uncertainty of 0\farcs4 in each coordinate (Fig. \ref{FCgrb121024A}). After the observations during the first night, imaging of the field of GRB 121024A continued on the second, third, fourth, sixteenth, and seventeenth night after the burst. The optical/NIR data were reduced using standard IRAF tasks \citep{1993tody, 2008kruehler}. The data were corrected for Galactic foreground reddening $E(B$-$V)$ = 0.09 mag \citep{2011ApJ...737..103S}, corresponding to an extinction of $A_{\rm{v}}=0.27$ mag for $R_{\rm{v}}=3.1$. The optical magnitudes were calibrated against secondary stars in the GRB field (Table \ref{refstars1}). On 2013 December 8 a Sloan Digital Sky Survey (SDSS) field \citep{2011ApJS..193...29A} at RA, Dec (J2000) = 04:59:42.0, -04:54:00 and the field of GRB 121024A were consecutively observed in photometric conditions. The calibration of the secondary stars was made against the corrected zeropoints of the GRB field based on the SDSS field. The NIR magnitudes were calibrated against the Two Micron All-Sky Survey (2MASS, \citealt{2006Skrutskie}) catalogue stars in the field of the GRB. 
\begin{table*}
\centering\setlength\tabcolsep{2.8pt}
\renewcommand*{\arraystretch}{1.0}
\caption{Secondary stars for photometric calibration. See Fig. \ref{FCgrb121024A} }
\label{refstars1}
\begin{tabular}{l*{9}{c}}
\toprule
\toprule
Star      & RA, Dec J(2000) & $g'(\rm{mag}_{\rm{AB}})$ &$r'(\rm{mag}_{\rm{AB}})$  &$i'(\rm{mag}_{\rm{AB}})$  & $z'(\rm{mag}_{\rm{AB}})$  & $J(\rm{mag}_{\rm{Vega}})$  & $H(\rm{mag}_{\rm{Vega}})$  & $K_{\rm{s}} (\rm{mag}_{\rm{Vega}})$ \\
\midrule
I   & 04:41:49.55, -12:16:47.2 & 19.96$\pm$0.05 & 18.75$\pm$0.05 & 18.21$\pm$0.06 & 17.92$\pm$0.06 & 16.73$\pm$0.06  & 15.96$\pm$0.07 & 15.88$\pm$0.08 \\
II   & 04:41:52.36, -12:16:49.9 & 17.83$\pm$0.05 & 17.13$\pm$0.05 & 16.84$ \pm$0.06 & 16.69$\pm$0.06 & 15.67$\pm$0.06  & 15.07$\pm$0.06  & 15.08$\pm$0.07  \\
III   & 04:41:55.40, -12:16:30.3& 20.73$ \pm$0.06 & 19.45$\pm$0.05 & 18.91$\pm$0.06 & 18.68$\pm$0.06 & 17.45$\pm$0.07  & 16.66$\pm$0.07 & --  \\
IV    & 04:41:47.91, -12:16:15.2 & 20.74$\pm$0.06& 19.23$\pm$0.05 & 18.44$\pm$0.06 & 18.04$\pm$0.06 &  16.79$\pm$0.06 & 16.07$\pm$0.06 & 15.96$\pm$0.06 \\
V   & 04:41:46.29, -12:17:57.5  & 17.76$ \pm$0.05 & 17.36$\pm$0.05 & 17.21$\pm$0.06 & 17.11$\pm$0.06  & 16.19$\pm$0.06 & 15.81$\pm$0.06 & 15.82$\pm$0.06 \\
VI  & 04:41:52.83, -12:18:46.8 & 20.57$\pm$0.05 &19.07$\pm$0.05 &18.43$\pm$0.06  &18.11$\pm$0.06 & 16.84$\pm$0.06 & 16.29$\pm$0.06 & -- \\
\bottomrule
\end{tabular}
\end{table*}

\subsection{APEX}
On 2012 October 24, we triggered an observation on the LABOCA bolometer camera \citep{refId0} \footnote{Based on observations collected during Max-Planck Society time at the Atacama Pathfinder Experiment (APEX) under proposal m-090.f-0005-2012.}. Two observations at a frequency of 345 GHz with a bandwidth of 60 GHz were performed. The first started $19.8$ ks after the GRB, the second $98.7$ ks after the GRB. On both days, the observations were taken in mapping mode and in on-off mode \citep{manualapex}. The data were reduced using the Bolometer Array analysis software \citep[BoA,][]{2012SPIE.8452E..1TSchuller}. All the subscans (ten per scan) were used. A clipping of 2 $\sigma$ was used to remove any background effects. The flux calibration was made using Jupiter for the focus, N2071IR as a secondary calibrator, and J0423-013 as a pointing source. There was no detection on either night, the upper limits are given in Table \ref{apexUL}.
\begin{table}[!ht]
\centering\small\setlength\tabcolsep{3pt}
\renewcommand*{\arraystretch}{1.0}
\caption{$1\sigma$ Upper limits of the on-off measurements using the LABOCA instrument on APEX.}
\label{apexUL}
\begin{tabular}{ c p{3.6cm} c }
\toprule
\toprule
Date & On+off time [UTC] & UL-Flux [mJy/beam]  \\  
\midrule
24-10-2012 & 08:22 - 09:20 & 3.6  \\
25-10-2012 & 06:16 - 06:38 & 10.4 \\
\midrule
Date & Mapping time [UTC] & UL-Flux [mJy/beam]  \\  
\midrule
24-10-2012 & 09:30-10:27, 10:39-11:00 & 9.0  \\
25-10-2012 & 08:52-09:29, 09:40-10:34, $\,$ 10:41-10:51 & 12.0 \\
\bottomrule
\end{tabular}
\end{table}

\subsection{Millimetre and radio observations}
In addition to the X-ray, GROND, and APEX data reported above, we also incorporated the following millimetre and radio observations reported in the literature in our SED analysis:

The Combined Array for Research in Millimetre-Wave Astronomy (CARMA) started observations of the field of GRB 121024A $\sim120.9$ ks after the BAT trigger at a mean frequency of $\sim85$ GHz (3mm) \citep{2012GCN..13900...1Z}. A mm counterpart was detected with a flux of $1.0\pm0.3$ mJy.

The Very Large Array (VLA) started observations of the field of GRB 121024A $\sim109.0$ ks after the trigger.  A radio counterpart with flux of $0.10\pm0.03$ mJy was detected at a frequency of $22$ GHz \citep{2012GCN..13903...1L}. 


\section{Phenomenological data analysis}
\label{analysis}

We start with a model-independent analysis of the data. The observed flux is described by $F\sim t^{-\alpha}\nu^{-\beta}$, with $\alpha$ and $\beta$ the temporal and spectral slope, respectively. First, we analyse the temporal evolution of the GRB 121024A afterglow. Using the X-ray and optical/NIR light curves, we measured the temporal slope $\alpha$ and obtained information about particular features such as flares, breaks in the light curve, flattening, or any behaviour different from that expected for a canonical afterglow light curve (LC) \citep{2006nousek,2006ApJ...642..354Z}. Then, we analysed the SED from X-ray to optical/NIR wavelengths at six different epochs. We obtained the spectral slope $\beta$ and checked for spectral evolution. Given that absorption and dust extinction only affect the data at X-ray and optical wavelengths, we used this SED analysis to derive the host X-ray absorbing column density ($N_{\rm{H}}^{\rm{host}}$), which is commonly quoted as an equivalent neutral hydrogen column density, and the host visual dust extinction along the GRB line of sight ($A^{\rm{host}}_{\rm{v}}$).

\subsection{Afterglow light-curve fitting}
\label{lcsec}
\begin{figure*}[ht!]
\centering
\includegraphics[width=\textwidth]{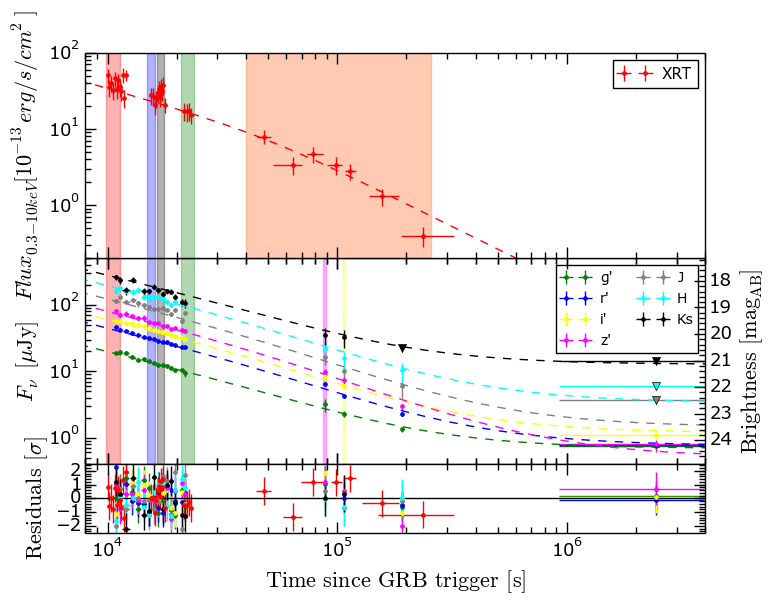}
\caption{Light curve of the afterglow of GRB 121024A. \textbf{Top}: XRT light curve from the XRT repository. \textbf{Bottom}: GROND light curve in $g'r'i'z'JHK_{\rm{s}}$. The best fit for the combined light curve (optical/NIR and X-ray data) is a smoothly broken power law with host contribution, shown with dashed lines. The epochs used for the spectral analysis are highlighted with the vertical bars. The break time $t_{\rm{b}}=49.8\pm5.1$ ks.}
\label{LCfit}
\end{figure*}

The temporal evolution of the X-ray afterglow of GRB 121024A\footnote{http://www.swift.ac.uk/xrt\_curves/536580} shows an initial steep decay with a temporal slope $\alpha = 3.6$, followed by a small flare at $\sim300$ s. For the present work, we only used the data after $10^4$ s, which is the start time of our GROND observations (Fig.\ref{LCfit}). We tried to fit two models: First, a simple power law with host contribution in the optical bands ($plh$) and slope decay $\alpha$. Second, a smoothly broken power law with constant host contribution ($brplh$) (Eq.\ref{smootheq}) \citep{1999AA...352L..26B}, with $\alpha_{\rm{pre}}$ and $\alpha_{\rm{post}}$ being the power-law slopes before and after the break, respectively, $sm$  the smoothness, and $t_{\rm{b}}$ the break time in the LC:
\begin{eqnarray}
\resizebox{0.9\hsize}{!}{$
F_{\nu}(t) = C \times \left\{ \left( \frac{t}{t_{\rm{b}}} \right)^{-\alpha_{\rm{pre}} \rm{sm}}+\left( \frac{t}{t_{\rm{b}}} \right)^{-\alpha_{\rm{post}} \rm{sm}}  \right\}^{-1/\rm{sm}} + host.$}
\label{smootheq}
\end{eqnarray}

The best fit to the X-ray light curve is a smoothly broken power
law with a statistical significance $\chi^{2}/\rm{d.o.f.}=51/44$ (simple power-law: $\chi^{2}/\rm{d.o.f.}=87/47$). The best-fit parameters are an initial decay with $\alpha_{\rm{pre}}=0.84\pm0.09$ and break time $t_{\rm{b}}^{\rm{xrt}}=32.5\pm16.1$ ks with $sm = 5.0\pm2.6$, followed by a steeper decay with $\alpha_{\rm{post}} =1.67\pm0.23$.\\

The optical/NIR light curves (Table \ref{grbmags}) are well fitted by both a $plh$ and a $brplh$ model. A $plh$ model gives $\chi^{2}/\rm{d.o.f.}=140/112$ and a decay slope $\alpha=1.07\pm0.02$, while a $brplh$ model gives $\chi^{2}/\rm{d.o.f.}=107/109$ and best-fitting parameters $\alpha_{\rm{pre}}=0.71\pm0.03$, $\alpha_{\rm{post}}=1.46\pm0.04$, break time $t_{\rm{b}}^{\rm{opt}}=31.4\pm9.4$ ks, and smoothness $2.7\pm1.1$.  Colour evolution in the optical bands is detected in the last epoch of our observations, which we associate with the increased contribution from the host galaxy. An F-test between the two model gives a null hypothesis probability of $1.86\times10^{-6}$. Therefore, we conclude that the $brplh$ profile describes the data in a better way. We used this profile for the following analysis. \\

Both the X-ray and optical/NIR light curves are best fitted by a broken power law with similar break times. We therefore tried a combined fit to both the XRT and GROND light curves to test whether the same model can describe both data sets, which would thus better constrain the best-fit parameters. The best-fit model provides a good fit ($\chi^2=157$ and $\rm{d.o.f.}=141$), with a pre-break temporal slope $\alpha_{\rm{pre}}=0.86\pm0.05$, post-break temporal slope $\alpha_{\rm{post}}=1.47\pm0.03$, smoothness $sm=1.7\pm0.3,$ and break time $t_{\rm{b}}=49.8\pm5.1$ ks. 

\begin{table*}
\centering\setlength\tabcolsep{3pt}
\renewcommand*{\arraystretch}{1.0}
\caption{Observed magnitudes of the GRB 121024A afterglow for the six highlighted epochs in Fig.\ref{LCfit}. The host contribution was subtracted. The magnitudes are not corrected for Galactic foreground extinction $A_{\rm{v}}^{Gal}=0.27$ mag.}
\label{grbmags}
\begin{tabular}{l*{9}{c}}
\toprule
\toprule
SED & mid-time [s] & $g'(\rm{m}_{\rm{AB}})$ &$r'(\rm{m}_{\rm{AB}})$  &$i'(\rm{m}_{\rm{AB}})$  & $z'(\rm{m}_{\rm{AB}})$  & $J(\rm{m}_{\rm{Vega}})$  & $H(\rm{m}_{\rm{Vega}})$  & $K_{\rm{s}} (\rm{m}_{\rm{Vega}})$ \\
\midrule
I & 11085 & 20.75$\pm$0.08 &19.82$\pm$0.06 & 19.53$\pm$0.06 & 19.24$\pm$0.05 & 18.68$\pm$0.10 & 18.31$\pm$0.11 & 17.91$\pm$0.13 \\
II & 15497 & 21.12$\pm$0.05 & 20.20$\pm$0.05 & 19.84$\pm$0.05 & 19.63$\pm$0.04 & 19.05$\pm$0.10 & 18.67$\pm$0.10 & 18.34$\pm$0.12  \\
III & 17006 &21.23$\pm$0.05 & 20.32$\pm$0.05 & 19.94$\pm$0.05 & 19.68$\pm$0.04 & 19.08$\pm$0.09 & 18.68$\pm$0.10 & 18.54$\pm$0.12 \\
IV & 21430 &21.48$\pm$0.27 & 20.54$\pm$0.06 & 20.20$\pm$0.06 & 19.95$\pm$0.09 & 19.43$\pm$0.10 & 18.89$\pm$0.12 & 18.81$\pm$0.15  \\
V & 88010 &22.89$\pm$0.24 & 22.03$\pm$0.09 & 21.74$\pm$0.10 & 21.49$\pm$0.14 & 21.01$\pm$0.26 & 20.76$\pm$0.31 & 20.41$\pm$0.31 \\
VI & 106998 &23.41$\pm$0.12 & 22.54$\pm$0.08 & 22.13$\pm$0.09 & 21.85$\pm$0.13 & 21.61$\pm$0.31 & 21.18$\pm$0.33 & 20.61$\pm$0.34 \\
\bottomrule
\end{tabular}
\end{table*}

\subsection{Afterglow SED fitting}
\label{sedsec}
\begin{figure}[!th]
\includegraphics[width=0.52\textwidth]{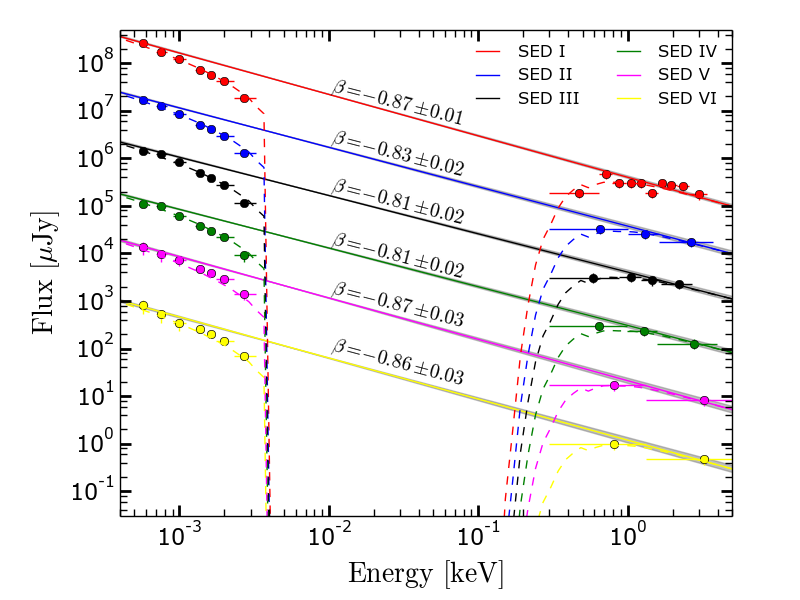}
\caption{Spectral energy distribution for the six SEDs highlighted in Fig. \ref{LCfit}. SEDs I - IV are from data before the observed break in the light curve. SEDs V $\text{and}$ VI are from data taken after the break. The SEDs are scaled with an arbitrary factor for clarity in the plot. The values of $\beta$ written above each line corresponds to the single power-law fit, where the slopes were left free to vary. The single power-law fit with a single tied slope has $\beta = 0.86 \pm 0.02$.}
\label{sed1fit}
\end{figure}

We analysed six different spectral epochs using XRT and GROND data, spanning the time interval $T_0+10$ ks to $T_0+240$ ks, four before the break time in the light curve and two after it (Fig. \ref{sed1fit}). The spectral analysis includes the effect of the dust and metal attenuation along the line of sight towards the source. For the last two SEDs, given the low signal-to-noise ratio at X-ray energies, we extracted the spectrum from the same time interval ($40$ ks - $240$ ks), during which time there was no evidence of spectral evolution within the X-ray energy range. We then renormalised the spectra so that they corresponded to the measured X-ray flux of the afterglow at the mid-time of the two corresponding SEDs (i.e. $t_{\rm{SED}}$ V = $88$ ks and $t_{\rm{SED}}$ VI = $107$ ks).\\

The SED analysis for all the six SEDs was performed simultaneously. The Galactic reddening was fixed to $E(B-V)=0.09$ mag, corresponding to an extinction of $A_{\rm{v}}^{Gal}=0.27$ mag \citep{2011ApJ...737..103S} for a Milky Way (MW) reddening law. The Galactic absorbing column density $N_{\rm{H}}^{\rm{Gal}}$ was fixed to $7.9\times10^{20}\,\rm{cm}^{-2}$ \citep{2005AA...440..775K}. The host magnitude was subtracted from the optical/NIR data, and the $g'$ band was not included in the fit because of a damped $\rm{Ly}\alpha$ system along the line of sight towards the GRB (DLA, \citealt{2015MNRAS.451..167F}). The values for the host extinction and absorption were tied between all the epochs, and the spectral slopes were left free to vary. A single power-law fit has a goodness of fit $\chi^2/d.o.f. = 28/46$ and all the spectral slopes values (see Fig. \ref{sed1fit}) are consistent within 1 $\sigma,$ confirming the lack of spectral evolution. A broken power-law fit either places the break outside the optical - X-ray frequency interval or fails to improve the fit when the break is forced to lie within this interval. In the latter case, the best-fit models have a goodness of fit $\chi^2/d.o.f = 32/40$\\

Given the lack of spectral evolution detected in our combined GROND/XRT light curve analysis out to $240$ ks, we fitted all six SEDs simultaneously with the same spectral model, with only the normalisation allowed to differ between epochs.  The best-fit results are given by a single power law with a spectral slope $\beta=0.86\pm0.02$ and goodness of fit $\chi^{2}/\rm{d.o.f.}=48/51$. The best-fit host dust extinction given by a Small Magellanic Cloud (SMC) reddening law \citep{1992ApJ...395..130P} is $A_{\rm{v}}^{\rm{host}} = 0.18 \pm 0.04$ mag, and the host galaxy X-ray absorbing column is $N_{\rm{H}}^{\rm{host}}=0.30^{+0.46}_{-0.29}\times10^{22}$ $\rm{cm}^{-2}$.  The lack of spectral evolution together with the achromatic break observed in the light curve rule out the movement of the cooling break through the observed wavelength range.\\

We extend the SED analysis in the following section with two additional epochs containing simultaneous observations with XRT, GROND, APEX, CARMA, and EVLA. The first SED at $t=21.9$ ks has GROND, XRT, and APEX data. The additional APEX upper limit requires a break between the APEX and NIR energies (see Fig. \ref{sedtotal2}). Then, we have a second SED at $t=109.0$ ks, with two additional measurements: CARMA and EVLA data points. The CARMA data point requires a break between the millimetre and the NIR bands, and the EVLA data point implies a break between the radio and the CARMA wavelength. Therefore at least two breaks in the broad-band spectrum of GRB 121024A are needed. These breaks are analysed in more detail in Sect. \ref{theory} in the context of the afterglow synchrotron spectrum model, where we use the constraints on $\beta$, $E(B-V)$ and $N_{\rm{H}}^{\rm{host}}$ found in his section.

\section{Physical parameters of the standard afterglow model}
\label{theory}
We now proceed with the derivation of the microphysical and dynamical parameters of the GRB afterglow, based on the standard afterglow model. In this model, the dominant emission is generally associated with synchrotron radiation from shock-accelerated electrons. These electrons are assumed to have a power-law energy distribution with slope $p$ and minimum energy $\gamma_{\rm{m}}$. The observed synchrotron spectrum is characterised by three main break frequencies ($\nu_{\rm{c}}$, $\nu_{\rm{m}}$, $\nu_{\rm{sa}}$) and a peak flux. The synchrotron injection frequency $\nu_{\rm{m}}$ is defined by $\gamma_{\rm{m}}$. The cooling frequency $\nu_{\rm{c}}$ is defined by the critical value $\gamma_{\rm{c}}$, above which electrons radiate their energy on timescales shorter than the explosion timescale. The self-absorption frequency $\nu_{\rm{sa}}$ marks the frequency below which the optical depth to synchrotron-self absorption is $>1$. In this model, two main cooling regimes are defined by the relative position of the break frequencies: a fast cooling regime where $\nu_{\rm{m}}>\nu_{\rm{c}}$ and most of the electron are cooling fast, and a slow cooling regime where $\nu_{\rm{m}}<\nu_{\rm{c}}$ and most of the accelerated electrons are cooling slowly \citep{1997meszarosapj,GS2002}. \\

The number of combinations of $\alpha$ and $\beta$ is limited when a specific dynamical model and the synchrotron spectrum are given. This gives rise to a unique set of relations between $\alpha$ and $\beta$ known as "closure relations" \citep{1997meszarosapj,1998ApJ...497L..17S,zhang_meszaros:2004}. These relations constrain the cooling regime, the circumburst environment, the jet geometry, and the electron energy distribution $p$. We follow two main steps to analyse the afterglow data: 

\begin{enumerate}
\item Spectral regime: The derivation of the $p$ value and identification of the external density profile depend on the power-law segment of the synchrotron spectrum containing the observing frequency. Using the closure relations \citep{2009racusin} together with the measured parameters for $\alpha$ and $\beta$, we find that the afterglow data can be described by two different spectral regimes (see Table \ref{ClosureTable}). In one case we have a spectral regime where $\nu_{\rm{c}} < \nu_{K_s}$ (i.e. below the $K_s$ band), and in the other case we have one where $\nu_{\rm{c}} > \nu_{\rm{xrt}}$ (i.e. above the XRT band). \\

\item Microphysical and dynamical parameters: We include the APEX, CARMA, and EVLA data in our analysis and fit the data using a single, a double, or a triple broken power-law model depending on each individual case (in the double and triple broken power-law fits, we only consider sharp breaks because the data at millimetre and radio frequencies are insufficient to constrain an additional free parameter i.e., smoothness). We use the standard formalism for a spherical blast wave propagating into an external cold medium during the slow cooling regime to derive all the micro-physical and dynamical parameters \citep{1997meszarosapj,GS2002}, and subsequently check for consistency with the slow or fast cooling transition times.
\end{enumerate}

Both spectral regimes, that is, $\nu_{\rm{c}} < \nu_{\rm{Ks}}$ $\nu_{\rm{c}} > \nu_{\rm{XRT}}$, are explained and analysed in detail in the following subsection. The former corresponds to the jet in the light curve being associated with a jet break without energy injection, and the latter corresponds to the jet in the light curve associated with either the end of energy injection into the outflow or with a jet break with an ongoing energy injection during the whole evolution of the afterglow (i.e. the ongoing energy injection is still visible until the last observations).
\begin{table*}[!ht]
\centering 
\renewcommand*{\arraystretch}{1.00}
\begin{threeparttable}[b]
\caption{Closure relations \tablefootmark{*}. $\beta = 0.86\pm0.02$ was use in the analysis. 
When determining the energy-injection parameter $q$, we use the measured $\alpha_{\rm{pre}} = 0.86\pm0.05$ and $\alpha_{\rm{post}} = 1.47\pm0.03$. The equations used for $q$ are for the case when $p>2$, for both spectral regimes. More details in \citealt{2009racusin}.}
\label{ClosureTable}
\begin{tabular}{ *{9}{c}} 
\toprule
\toprule

& &   \multicolumn{3}{c}{Instantaneous injection} &  \multicolumn{3}{c}{Energy injection} &  \multirow{2}{*}{Section\tablefootmark{**}} \\

& &  \multicolumn{3}{  c }{$\alpha (\beta)$} &  \multicolumn{3}{ c }{$q (\beta,\alpha)$}\\

\midrule
& & \multicolumn{7}{ c }{\multirow{1}{*}{Spherical outflow}}\\
\midrule
\midrule

\multirow{2}{*}{$ \nu_{\rm{c/m}} < \nu$ \tablefootmark{a}} & ISM  & $\frac{3\beta+5}{8}$  &  $=$ & $\textbf{0.95}\pm\textbf{0.01}$ & $\frac{2(1+\alpha-\beta)}{\beta+1}$ & $=$ & $1.07\pm0.05$ & \multirow{2}{*}{\ref{jetbreak}}\\

& Wind  & $\frac{\beta+3}{4}$ & $=$ & $\textbf{0.96}\pm\textbf{0.01}$ & $\frac{2(1+\alpha-\beta)}{\beta+1}$ & $=$ & $1.07\pm0.05$  &  \\
\midrule

 \multirow{2}{*}{$\nu_{\rm{m}} < \nu < \nu_{\rm{c}} $ \tablefootmark{b}} & ISM & $\frac{3\beta}{2}$ & $=$ & $ 1.29\pm0.03$ & $\frac{2(1+\alpha-\beta)}{\beta+2}$ & $=$ & $ \textbf{0.69}\pm\textbf{0.04}$  & \multirow{2}{*}{\ref{einj}} \\
 & Wind & $\frac{3\beta + 1}{2}$ & $=$ & $1.79\pm0.03 $ & $\frac{2(\alpha-\beta)}{\beta+1}$ & $=$ & $-0.01\pm0.05$ &  \\


\midrule
& & \multicolumn{7}{ c }{\multirow{1}{*}{Uniform non-spreading jet}} \\
\midrule
\midrule
\multirow{2}{*}{$ \nu_{\rm{c/m}} < \nu$ \tablefootmark{a}} & ISM &  $\frac{3\beta+11}{8}$  &  $=$ & $1.70\pm0.01$ & $\frac{2(1+2\alpha-2\beta)}{3+2\beta}$ & $=$ & $ 0.95\pm0.05$ & \multirow{2}{*}{\ref{jetbreak}}\\
& Wind & $\frac{\beta+5}{4}$ & $=$ & $\textbf{1.47}\pm\textbf{0.01}$ & $\frac{2(1+\alpha-\beta)}{2+\beta}$ & $=$ & $1.13\pm0.04$  \\
\midrule
 \multirow{2}{*}{$\nu_{\rm{m}} < \nu < \nu_{\rm{c}} $ \tablefootmark{b}}  & ISM & $\frac{6\beta+3}{4}$ & $=$ & $ 1.64\pm0.01$  & $\frac{2(1+2\alpha-2\beta)}{5+2\beta}$ & $=$ & $ \textbf{0.65}\pm\textbf{0.03}$ &  \multirow{2}{*}{\ref{Einj2}} \\
 & Wind & $\frac{3\beta + 2}{2}$ &$=$ &  $1.84\pm0.01 $ & $\frac{2(\alpha-\beta)}{2+\beta}$ &  $=$ &$0.43\pm0.04$ \\

\midrule
& & \multicolumn{7}{ c }{\multirow{1}{*}{Uniform spreading jet}}\\
\midrule
\midrule
 $ \nu_{\rm{c/m}} < \nu$ \tablefootmark{a}& ISM/Wind & $\frac{\beta+3}{2}$  &  $=$ & $1.93\pm0.01$ &  $\frac{2+3\alpha-4\beta}{2(\beta+1)}$ & $=$ & $0.79\pm0.05$ & \ref{Einj2}\\
 \midrule
$\nu_{\rm{m}} < \nu < \nu_{\rm{c}} $ \tablefootmark{b} &  ISM/Wind & $2\beta+1$ & $=$ & $ 2.72\pm0.04$  & $\frac{1+3\alpha-4\beta}{2(\beta+2)}$  & $=$ &  $ 0.35\pm0.03$ & \ref{Einj2}\\

\bottomrule
\end{tabular}

  \begin{tablenotes}\footnotesize
   \item[*] When $\nu_c < \nu < \nu_m$ $\beta = 0.5$ and it does not depend on p or $\alpha$. We did not include this scenario as it is not compatible with our data at any time.
   \item[**] Details on the results and implications of the closure relations are discussed in the outlined section.
    \item[a] For $ \nu_{\rm{c}} > \nu$, $p=2\beta$. When $\beta = 0.86\pm0.02$ we have $p=1.73\pm0.03$ ($1 < p < 2$).
    \item[b] For $\nu_{\rm{m}} < \nu < \nu_{\rm{c}} $. When $\beta = 0.86\pm0.02$ we have $p=2.73\pm0.03$ ($ p > 2$).

  \end{tablenotes}
 \end{threeparttable}
\end{table*}

\subsection{$\nu_{\rm{c}} < \nu_{K_s}$: Jet break.}
\label{jetbreak}

Using the closure relations for a decelerating spherical blast wave, we find that the measured temporal slope before the break in the light curve is consistent with $\nu_{\rm{c}} < \nu_{K_s}$ for both ISM and wind environments. This implies $p=1.73\pm0.03$, as $\beta=p/2$. The only plausible scenario consistent with the measured $\alpha_{\rm{post}}$ and $\beta$ corresponds to a non-spreading uniform jet propagating into a wind environment. We therefore associate the achromatic break observed in the light curve with a jet break \citep{1999ApJ...525..737R,2014Natur.509..201W}. \\

We proceed by including the post-break sub-mm and radio data in our analysis. The first broadband SED contains GROND, XRT, and APEX data. The best fit to this is a broken power law with both Galactic and host extinction and absorption, with $\chi^2/\rm{d.o.f.}=3.6/5$ (see Table \ref{SEDresultsfreq}). The measured value of $\nu_{\rm{c}}=1.5\times10^{12}$ Hz is a lower limit because the APEX measurement is an upper limit. The second broadband SED contains XRT, GROND, CARMA, and EVLA detections. Two possible spectral sub-regimes in the slow cooling phase give a good fit to the data: The cooling regime where $\nu_{\rm{sa}}$ < $\nu_{\rm{m}} < \nu_{\rm{c}}$, and the one where  $\nu_{\rm{m}}$ < $\nu_{\rm{sa}} < \nu_{\rm{c}}$. Because
there are only a few data points at radio wavelengths, it is difficult to distinguish between these two cooling regimes. Therefore, we analysed both cases. 

\begin{table}[!ht]
\centering\setlength\tabcolsep{2pt}
\renewcommand*{\arraystretch}{1.1}
\caption{Results from SED fits for both a jet-break model with $\nu_{\rm{sa}}$ < $\nu_{\rm{m}}$ and $\nu_{\rm{m}}$ < $\nu_{\rm{sa}}$, and for an energy-injection model.}
\label{SEDresultsfreq}
\begin{tabular}{c c c c c}
\hline
\hline
 $\nu$ & Time & Jet break & Jet break & Energy \\
$[\rm{Hz}]$ & [ks] & $\nu_{\rm{sa}}$<$\nu_{\rm{m}}$ & $\nu_{\rm{m}}$<$\nu_{\rm{sa}}$ & injection \\
\hline 
\multirow{2}{*}{$\nu_{\rm{c}}$}& $26$ & $>1.5\times10^{12}$ & $>1.5\times10^{12}$ & $>1.2\times10^{18}$ \\
& $109$ & $1.9^{+5.2}_ {-0.4}\times10^{12}$ & $3.9^{+3.2}_{-2.4}\times10^{12}$ & $ > 1.2\times10^{18}$  \\ 
\hline
\multirow{2}{*}{$\nu_{\rm{m}}$} & $26$ & -- & --  & $<1.4\times10^{14}$  \\
& $109$ & $1.3^{+1.3}_{-0.3} \times10^{11}$ & $<2.2\times10^{10}$  & $ 5.1^{+1.9}_ {-0.6}\times10^{11}$ \\
\hline
\multirow{2}{*}{$\nu_{\rm{sa}}$} & $26$ & -- & -- & --  \\
& $109$ & $8.3^{+1.7}_{-1.6} \times10^{10}$ & $ 7.4^{+2.6}_{-0.7}\times10^{10}$ & $7.4^{+0.2}_ {-1.6}\times10^{10}$  \\
\hline 
\end{tabular}
\end{table}


When $1<p<2$ (i.e. a hard electron spectrum), there is more energy-per-decade in high-energy electrons. This distribution has important implications for the analysis of the physics in the shock region, specifically requiring an additional high-energy cut-off in the electron population. We based our analysis on the assumption of a proportionality between $\gamma_{\rm{m}}$ and $\gamma$, where $\gamma$ is the Lorentz factor of the shocked fluid. This implies that $\gamma_m$ is proportional to local temperature, which is physically plausible since the non-thermal population is presumably accelerated out of a Maxwellian population. The upper cut-off in the electron distribution can be assumed to lie beyond the X-ray band and does not need to be accounted for explicitly. $\epsilon_e$ can no longer be interpreted as the fraction of energy in accelerated electrons. Instead, it becomes a scale factor between $\gamma$ and $\gamma_m$, according to $\gamma_{\rm{m}}=\rm{K}\times \gamma$ with $\rm{K}=\bar{\epsilon}_{\rm{e}}\times m_{\rm{p}}/m_{\rm{e}}$ \citep{1997meszarosapj}. We followed the formalism used by \citet{GS2002}, who derived the flux equation using a full fluid profile for the blast wave \citep{BM1976} and took the line-of-sight effect and the cooling times of the individual electrons into account.\\

\begin{figure}[h!]
\centering
\subfloat[Jet break: $\nu_{\rm{sa}} < \nu_{\rm{m}} < \nu_{\rm{c}}$]{\includegraphics[width=0.44\textwidth]{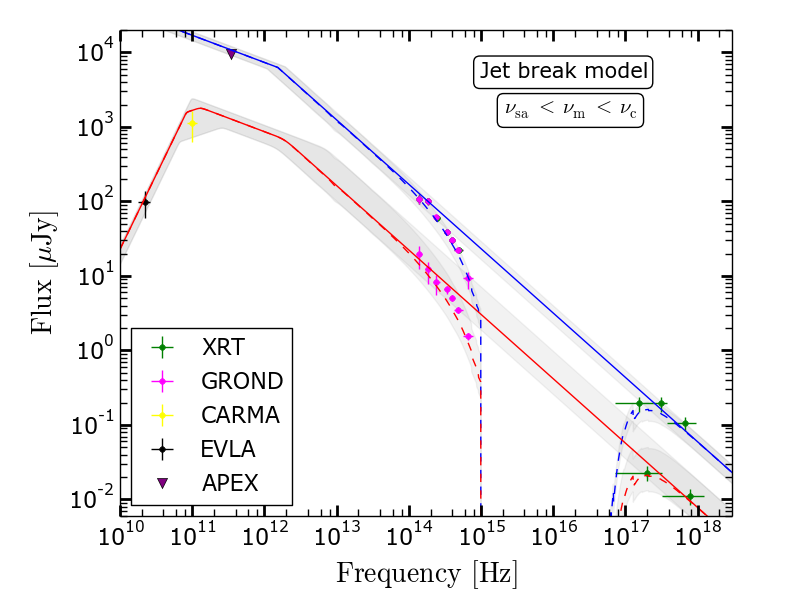}\label{break1}} \\
\subfloat[Jet break: $\nu_{\rm{m}} < \nu_{\rm{sa}} < \nu_{\rm{c}}$]{\includegraphics[width=0.44\textwidth]{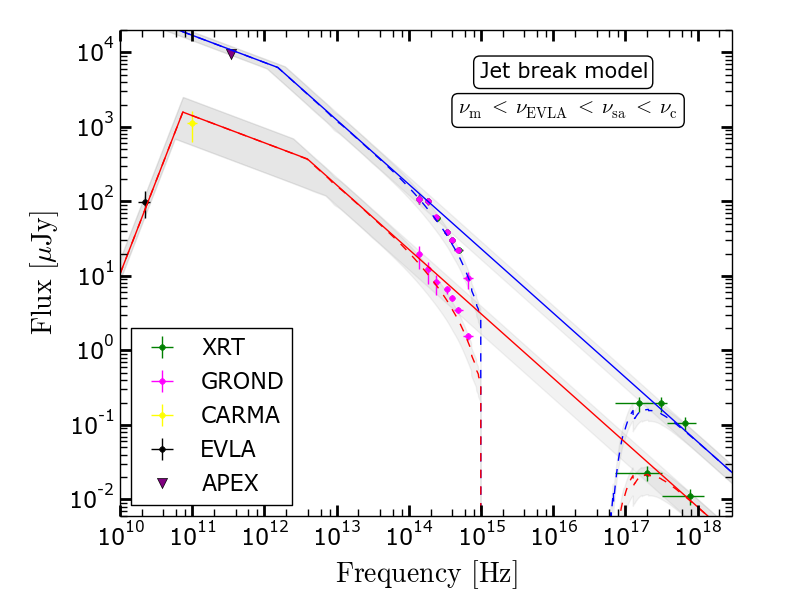}\label{break2}}\\
\subfloat[Energy injection]{\includegraphics[width=0.44\textwidth]{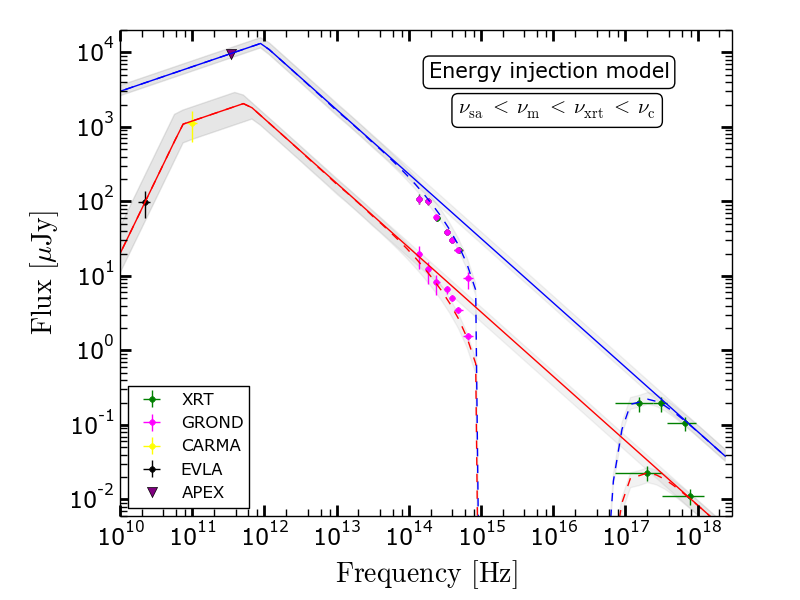}\label{einjection}}
\caption{Broadband SEDs of the afterglow of GRB 121024A from the radio to the X-ray regime for the three models described in Sect. \ref{theory}. Blue line: SED at $t=21.9$ ks. Red line: SED at $t=109.0$ ks. The dashed lines represent the absorbed model, the solid lines the unabsorbed model. The grey-shaded regions corresponds to the $1\sigma$ limits of the model.}
\label{sedtotal2}
\end{figure}

Based on the values for the break frequencies presented in Table \ref{SEDresultsfreq} for both spectral regimes, $\nu_{\rm{sa}}$ < $\nu_{\rm{m}}$ (see Fig. \ref{break1}) and $\nu_{\rm{m}}$ < $\nu_{\rm{sa}}$ (see Fig. \ref{break2}), we derived the microphysical and dynamical parameters. The results are reported in Table \ref{paramstable} and are used to calculate the transition times between the spectral regimes. First, the transition from fast to slow cooling. This corresponds to $t_{\nu_c = \nu_m}\sim2.8\times10^4$ s and $t_{\nu_{c} = \nu_m}\sim2.6\times10^{3}$ s for  $\nu_{\rm{m}}$ > $\nu_{\rm{sa}}$ and  $\nu_{\rm{m}}$ < $\nu_{\rm{sa}}$, respectively. In both cases, it is before the time of the analysed SED at $t = 109$ s, confirming the slow cooling assumption. Second, the transition from optically thin to optically thick, that is, when $\nu_m$ goes below $\nu_{\rm{sa}}$. This occurs at $t_{\nu_{sa} = \nu_m}\sim1.8\times10^5$ s when  $\nu_{\rm{m}}$ > $\nu_{\rm{sa}}$ and at $t_{\nu_{sa} = \nu_m}\sim1.07\times10^5$ s when  $\nu_{\rm{m}}$ < $\nu_{\rm{sa}}$. \\

\subsection{$\nu_{\rm{c}} > \nu_{\rm{xrt}}$ : Energy injection.}
\label{einj}

The closure relations (Table. \ref{ClosureTable}) and the possible spectral break positions resulting from fitting synchrotron spectra to the SED allow for an alternative scenario, where $\nu_c > \nu_{XRT}$ and p $>$ 2. In this case, the break between the mm and NIR wavelength corresponds to $\nu_{\rm{m}}$ and the break in the light curve is associated with the end of the ongoing energy-injection phase. Smooth energy injection into the ejecta can result from slower shells with a range of velocities that
catch up with each other or from a long-term engine luminosity. In the latter case, the energy-injection parameter $q$ is defined by $\rm{L} = \rm{L}_0 (t/t_{\rm{b}})^{-q}$. Using the flux and frequency equations for radial flow from \citet{2009MNRAS.394.2164V} and \citealt{2012MNRAS.427.1329L}, we derived the closure relation for a general density profile with an arbitrary $k$ during the deceleration stage following energy injection ($k = \frac{4(2\alpha-3\beta)}{1+2\alpha-3\beta}$ for $\nu_{\rm{m}} < \nu < \nu_{\rm{c}}$). The best-fit results for $\alpha_{\rm{post}}$ and $\beta$ then imply $k =1.05\pm0.23$. \\

During the energy-injection phase, a forward-reverse shock system is set up in the flow. Using the flux equation describing the energy-injection phase from \citet{2014MNRAS.442.3495V}, we have a given relation between $\alpha$, $\beta$, $k$ and $q$. If the emission is dominated by that from the forward shock (FS):

\begin{equation}
\label{closureFS}
FS ~ : ~ q = \frac{8-2\alpha(-4+k)+2\beta(-4+k)-4k}{3k-8+\beta(k-4)},
\end{equation}

and the following relation if the reverse shock (RS) emission is dominant:
\begin{equation}
\label{closureRS}
RS ~ : ~ q = \frac{4+8\alpha-2(1+\alpha+\beta)k}{(3+\beta)k-10}.
\end{equation}

The values for $\alpha_{\rm{pre}}$ and $\beta$ derived in Sect. \ref{lcsec} imply $q = 0.52 \pm 0.07$ for dominant FS emission and $q=0.88\pm0.09$ for dominant RS emission. These $q$ values (as well as the pre-break temporal slopes) are consistent with those determined for \textit{Swift} samples (e.g. \citealt{2009racusin,2009MNRAS.397.1177Evans,2013MNRAS.428..729M}). If we fix $k = 2$, we obtain $q \simeq 0$ for both RS and FS, consistent with predictions for a magnetar model \citep{1998PhRvL..81.4301D, zhang_meszaros:2004}. For an ISM density profile, $q = 0.69\pm0.04$ for FS emission and $q = 1.09 \pm 0.03$ for RS emission. RS emission can therefore not be dominant because $q>1$ implies that the energy injection decays too rapidly to sustain a plateau.\\

After the energy-injection phase, only a decelerating forward shock remains, and a standard afterglow emission model can be applied. We therefore proceeded with the analysis of the final SED at $t = 109s$, which contains EVLA, CARMA, GROND, and XRT data. The best-fit profile is a sharp double broken power law with $\chi^2/\rm{d.o.f.} = 8.50/8$ (Fig. \ref{einjection}). The critical values reported in Table  \ref{SEDresultsfreq} were used to derive the micro-physical parameters after the energy-injection phase (see Table \ref{paramstable}). In this scenario, $\nu_{\rm{c}}$ cannot be measured and we can only place a lower limit. The $k$ value lies just between the expected values for ISM and wind environments, and therefore we determined the values for both wind and ISM environments using \cite{GS2002} and for $k=1.05$ using \cite{2009MNRAS.394.2164V} and \cite{2012MNRAS.427.1329L}. \\

\subsection{$\nu_{\rm{c}} > \nu_{\rm{xrt}}$ : Energy injection and jet break.}
\label{Einj2}

Now, we analyse the afterglow parameters assuming prolonged energy injection at all times, both before and after the break in the light curve, and the break in the light curve is associated with a jet break. As shown in Sect. \ref{einj}, RS emission is not dominant before the break, and therefore we assumed only a dominant FS emission. To avoid too many free parameters, we restricted the study to ISM and wind density profiles. The analysis for these two medium profiles for the ongoing energy-injection phase before the break in the light curve is presented in Sect. (\ref{einj}). Here, we used $\alpha_{\rm{post}}$ to derive the $q$ values after the break and, assuming $q$  does not evolve, we compared these post-break values with the pre-break values to check if it is possible to have ongoing energy injection together with a jet break. We analysed two cases for the post-jet-break evolution: a sideways spreading jet and a non-spreading jet. For the former case  $q=0.35\pm0.03$, inconsistent with the $q$ value before the jet break. In addition to this inconsistency, if the energy is continuously injected within $\theta_0$, while the front of the jet begins to spread, the homogeneous shell approximation leading to the closure relations used here is no longer valid. On balance, the bulk of the energy will remain confined to $\theta_0$ (see discussion in \citealt{2014MNRAS.442.3495V}), and then the non-spreading jet approximation is favoured. The non-spreading case gives $q=0.65\pm0.03$ for an ISM density profile and $q=0.43\pm0.03$. Therefore, only the ISM density profile is consistent with the pre-break $q$ value, and the prolonged energy injection would only be possible if the observed jet break is due to geometrical effects alone. \\

We analysed our last two SEDs including radio, submm, NIR, optical, and X-ray data. We used the flux and break frequency equations for energy injection presented in \citet{2014MNRAS.442.3495V} together with equations for $\nu_{sa}$ (van Eerten in prep.) to obtain the model parameter values presented in Table \ref{paramstable}. As in the previous case, where energy injection was only operating before the break in the light curve, some unphysical values for the parameters are found. The main problems are $\epsilon_{e} > 7.6$, when it should not be greater than unity, and the value for the density $\sim 10^{7}$ $\rm{cm}^{-3}$, instead of being of order unity as expected.

\begin{table*}[!ht]
\centering\setlength\tabcolsep{5.0pt}
\renewcommand*{\arraystretch}{1.2}
\caption{$\gamma_{\rm{m}}$,$\bar{\epsilon}_{\rm{e}}$, $\epsilon_{\rm{B}}$, $E_{\rm{iso}}$, $n$ and $\theta_0$ for the models described in Sects. \ref{jetbreak}, \ref{einj}, and \ref{Einj2}. $\bar{\epsilon}_{\rm{e}} = \epsilon_{\rm{e}} \times (|p-2|)/(p-1)$ and $E_{\rm{iso},52}=E_{\rm{iso}}/10^{52}$. The half-opening angle is derived using Eq.(4) from \cite{2005ApJ...618..413G}. $n = \rm{A}r^{-2}$ with $\rm{A}=\dot{M}/4\pi \rm{v}_{\rm{w}}=5\times10^{11}\rm{A}_{*}$ g cm$^{-1}$ \citep{2000ApJ...536..195C}. For $k = 2$ we report the density in terms of $A_{*}$. For $k=0$ and $k=1.05$ we report the number density $n_0$ in units of cm$^{-3}$. In the special case of $k=1.05$ we use a reference distance of $r=10^{17}$ cm.}
\label{paramstable}
\begin{tabular}{c|c|c|c|c|c|c|c}
\toprule
\toprule
\multicolumn{1}{c}{} & $\gamma_{\rm{m}}$ & $\bar{\epsilon}_{\rm{e}}$& $\epsilon_{\rm{B}}$ & $A_{*}$ , $n_0$ & $\theta_0$ [rad]  & $E_{\rm{iso},52}$[erg] & $\,$ $\eta$ $\,$\\
\midrule
 \multicolumn{8}{c}{Jet break, $\,\,\,\,\,\,$ $p=1.73\pm0.03$, $\,\,\,\,\,\,$ $\nu_{\rm{c}} < \nu_{\rm{K_s}}$}  \\
\midrule
$\nu_{\rm{sa}} < \nu_{\rm{m}}$ & $102.7^{+139.6}_{-54.2}$  & $2.09^{+2.86}_{-1.05}\cdot10^{-2}$ & $2.11^{+2.49}_{-0.91}\cdot10^{-2}$ & $1.41^{+4.01}_{-1.47}$ & $0.32^{+0.07}_{-0.02}$ & $0.15^{+0.07}_{-0.03}$ & $98^{+2}_{-3}\%$\\
$\nu_{\rm{m}} < \nu_{\rm{sa}}$  & < $11.2$  & < $9.31\cdot10^{-4}$ & < $7.87\cdot10^{-2}$ & > $0.78$ & > $0.13$ & $>2.94$ & $<74\%$\\
\midrule
\midrule
\multicolumn{8}{c}{Energy injection until $t_b$ in the light curve, $\,\,\,\,\,\,$ $p=2.73\pm0.03$, $\,\,\,\,\,\,$  $\nu_{\rm{c}} > \nu_{\rm{xrt}}$}\\
\midrule
$k =2$  & > $2.01\cdot10^{3}$  & > $1.10$ & < $6.64\cdot10^{-10}$ & > $1.23\cdot10^{3}$ & >$0.85$ & > $2.36$ &  $<78\%$\\
$k =1.05$ & > $1.4\cdot10^{3}$  & > $0.76$ & < $2.1\cdot10^{-9}$ & > $4.3\cdot10^{5}$ &>$0.8$ & > $3.4$ & $<71\%$ \\
$k =0$  & > $1.11\cdot10^{3}$  & > $0.75$ & < $2.25\cdot10^{-9}$ & > $1.21\cdot10^{7}$ &>$0.77$ & > $3.67$  &  $<69\%$\\
\midrule
\multicolumn{8}{c}{Energy injection scenario with jet break, $\,\,\,\,\,\,$ $p=2.73\pm0.03$, $\,\,\,\,\,\,$ $\nu_{\rm{c}} > \nu_{\rm{xrt}}$}\\
\midrule
$k =0$  & > $1.6\cdot10^{4}$  & > $7.6$ & < $3.9\cdot10^{-8}$ & > $1.29\cdot10^{7}$ & > $1.21\cdot10^{-2}$ & > $0.16$  &  $<98\%$\\
\bottomrule
\end{tabular}
\end{table*}

\section{Discussion}
\label{dis}
In the previous sections we have presented a detailed analysis of the afterglow observations and derived values for the microphysical and dynamical parameters. Here we compare the derived values in the different scenarios. We discuss the positive and negative aspects of each model in the framework of the standard afterglow model.

\subsection{Jet break without energy injection}

This scenario requires three main features: First, the cooling break must be at around a few times $10^{12}$ Hz at $> T_0 + 109$ ks. Although a value for $\nu_{\rm{c}}$ this low has been seen before (i.e. GRB 060418, \citealt{Cenko2010}), in more than $95\%$ of a combined GROND-XRT sample, $\nu_{\rm{c}}$ was detected above the optical frequencies \citep{2011AA...526A..30G}. Second, the closure relations require that the jet does not spread out sideways following the break time. The jet has to remain in this non-spreading state at least until $ \text{about one}$ day after the jet break because no spectral evolution is detected so far in the observations. This behaviour is at odds with findings from theoretical \citep{2012MNRAS.421..570G} and numerical \citep{2010ApJ...722..235V,2012ApJ...751...57D,2012ApJ...751..155V} studies of afterglow jets. Third, a very hard electron spectrum ($p<2$) with $p=1.73$ is required, as are additional assumptions about the minimal Lorentz factor. Although this is significantly lower than the value of 2.3 expected from Fermi acceleration theory (e.g. \citealt{2000ApJ...542..235K,achterberg2001}), it is within the average range of values $1.5-3.0$ observed in previous GRB afterglow studies (\citealt{2010ApJ...716L.135C}). \\ 

Two different spectral sub-regimes were presented in Sect. \ref{jetbreak}, either with $\nu_{\rm{sa}} < \nu_{\rm{m}}$ or with $\nu_{\rm{m}} < \nu_{\rm{sa}}$. Here the main assumption is $\gamma_{\rm{m}}\sim\gamma$. For both spectral sub-regimes, the derived values for $\epsilon_{\rm{B}}$ (Table \ref{paramstable}) are in the same range as previous measurements reported in the literature, and the values of $\theta_0$ are consistent with a collimated outflow ($0.1-0.3$ rad). The values for the circumburst density therefore agree with the collapsar model and a Wolf-Rayet star as possible progenitor, with mass-loss rates of $\sim 1.4\times10^{-5} \, M_{\odot} \, \rm{yr}^{-1}$ when $\nu_{\rm{sa}} < \nu_{\rm{m}}$ and $> 7.8\times10^{-6} \, M_{\odot} \, \rm{yr}^{-1}$ when $\nu_{\rm{m}} < \nu_{\rm{sa}}$, for a wind velocity $v = 1000$ km s$^{-1}$ \citep{1999ApJ...520L..29C,2000ApJ...536..195C}. The efficiency\footnote{Efficiency of the conversion of the kinetic energy in the outflow to gamma-rays during the prompt emission $\eta=\rm{E}_{\rm{iso},\gamma}/(\rm{E}_{\rm{iso},\gamma}+\rm{E}_{\rm{iso}})$. $E_{\rm{iso},\gamma}$ is the isotropic energy released in the prompt gamma-ray emission. In this case $E_{\rm{iso},\gamma} = 8.4^{+2.6}_{-2.2}\times10^{52}$ erg \citep{2007butler} (http://butler.lab.asu.edu/Swift/index.html). It is calculated using $E_{\rm{iso},\gamma} = 4\pi d_{\rm{L}}^2\rm{F}/(1+z)$, where F is the fluence in the gamma-ray band. BAT: from $15-150$ keV in the observer-frame. $E_{\rm{iso}}$: energy range $1-10^{4}$ keV in the rest frame.}  requirements are extremely high. For $\nu_{\rm{sa}} < \nu_{\rm{m}}$ the measured $E_{iso}$ implies an efficiency of $\eta \sim 98\%$ and for $\nu_{\rm{m}} < \nu_{\rm{sa}}$ the efficiency is $\eta<74\%$. These two efficiency values are much higher than expected in the standard fireball shock model, for which an efficiency of $\eta < 10\%$ is predicted \citep{1997ApJ...490...92K,1998MNRAS.296..275D,1999ApJ...523L.113K,2006MNRAS.370.1946G,Cenko2011}. \\

As a final verification of this model, we applied a condition from \citet{GS2002} to the evolution of the afterglow spectrum in a wind environment for a given set of microphysical and environmental parameters. This states that if $A_{*}\bar{\epsilon_{\rm{e}}}^{-1}E_{\rm{iso,52}}^{-3/7}\epsilon_{\rm{B}}^{2/7} > 100,$ the afterglow spectrum evolves from fast to slow cooling, where in the slow cooling phase, initially $\nu_{\rm{sa}} < \nu_{\rm{m}} < \nu_{\rm{c}}$, but eventually $\nu_{\rm{m}} < \nu_{\rm{sa}} < \nu_{\rm{c}}$. If $A_{*}\bar{\epsilon_{\rm{e}}}^{-1}E_{\rm{iso,52}}^{-3/7}\epsilon_{\rm{B}}^{2/7} < 100,$ the afterglow spectrum only goes through one spectral regime in the slow cooling phase where $\nu_{\rm{m}} < \nu_{\rm{sa}} < \nu_{\rm{c}}$. In this latter scenario we are never in the regime where $\nu_{\rm{sa}} < \nu_{\rm{m}}$ during the slow cooling phase. In our jet-break model where the spectral regime is $\nu_{\rm{sa}} < \nu_{\rm{m}}$, we therefore require that the derived micro-physical and dynamical parameters give $A_{*}\bar{\epsilon_{\rm{e}}}^{-1}E_{\rm{iso,52}}^{-3/7}\epsilon_{\rm{B}}^{2/7} > 100$. However, we find that our best-fit values presented in Table \ref{paramstable} for $\nu_{\rm{sa}} < \nu_{\rm{m}}$ give 52, inconsistent with the condition stated above, and therefore this regime can be ruled out. The favoured regime is then a slow cooling phase where $\nu_{\rm{m}} < \nu_{\rm{sa}}$, where our best-fit parameters give the value 257.

\subsection{Energy injection}
\label{einjdis}

According to the shape of the spectrum and the closure relations, it is also possible to have $\nu_{\rm{c}} > \nu_{\rm{XRT}}$, implying an energy-injection model. In the energy-injection scenario, both with and without a jet break, our best-fit values for $q$ are consistent with $q\sim0.5$, corresponding to smooth energy injection, which has been observed in several other cases (e.g. \citealt{2006ApJ...642..354Z}), and the hard electron spectrum is not required anymore since we now have $p=2\beta+1=2.73$. However, further problems with the other afterglow parameters are found.\\

The energy-injection scenario together with a jet break in an ISM external medium and without a jet break in a wind-like ($k=2$) external medium can be ruled out because $\epsilon_{\rm{e}} > 1$, therefore, this scenario is physically not meaningful. There are two other scenarios to be analysed then: the energy-injection scenario without a jet break in an ISM medium and with a general density profile with slow $k=1.05$. The analysis below focuses on these two cases.\\

In relation to the micro-physical parameters, the derived values for $\epsilon_{\rm{B}}$ differ from previous observations, but agree with theoretical predictions. In the former case $\epsilon_{\rm{B}}$ is more than four orders of magnitude ($<10^{-9}$) smaller than the average measured values from previous studies (e.g. \citealt{Panaitescu2002,2003ApJ...597..459Y,Panaitescu2005,Cenko2010}). In the latter case, the value $\epsilon_{\rm{B}} < 10^{-9}$ for an ISM density profile is consistent with expected values from shock compression of the seed magnetic field (B$_{0}$$\sim\mu$G) in the surrounding medium \citep{2009MNRAS.400L..75K,santana2014}, and no further amplification or additional magnetic field would be required in the shock region. On the other hand, the derived value for $\bar{\epsilon}_{\rm{e}}$ is $< 0.75$ consistent with theory as $\epsilon_{\rm{e}} < 1$ but higher than the average of observations where $\epsilon_{\rm{e}} \sim 0.2$ \citep{santana2014}.\\

Furthermore, in the case where $\epsilon_{\rm{B}}\ll\epsilon_{\rm{e}}$, as implied by our analysis, we would  expect there to be a contribution to the cooling of electrons from inverse Compton (IC) scattering processes  \citep{2000ApJ...543...66P,2001ApJ...548..787S}. The IC emission will mainly affect the cooling frequency in the slow cooling regime. The final value of $\nu_c$ is expected to be lower than the value with synchrotron cooling alone. The IC contribution to the total observed afterglow emission can be included using the Compton parameter defined as $Y = \eta_c \epsilon_{\rm{e}}/\epsilon_{\rm{B}}$, where $\eta_c = (\gamma_c/\gamma_m)^{2-p}$ for the slow cooling regime (for more details see \citealt{2001ApJ...548..787S}). With this parameter the cooling frequency will be lowered by a factor of $(1+Y)^2$. A constraint $C$ can be derived to test whether the IC contribution is important during the evolution of the observed emission from the afterglow, or if it can be neglected (see Eq. (4.9) of \citealt{2001ApJ...548..787S}). This constraint depends only on the observational quantities (break frequencies and peak flux) of the afterglow and is independent of the theoretical afterglow parameters. It can be expressed in terms of $Y$ as $C = Y/(1+Y)^{2}$ and has a consistent solution for Y only if $C < 1/4$. Using our measurements, we find $C \sim 10^6$ for ISM. This result indicates that the IC component is not a relevant contribution for this afterglow, contrary to the theoretical expectation when $\epsilon_{\rm{B}}/\epsilon_{\rm{e}} << 1$ and the energy-injection scenario is not favoured.\\

The lower limits derived for the density are 2 ($k=1.1$) and 4 ($k=0$) orders of magnitude higher than previous density measurements for bursts with similar isotropic energies ($E_{\rm{iso}} \sim 10^{52}$ erg) between $10^{-2}$ and $10^{3}$ for constant density circumburst profiles ($n$) (e.g. see Fig. 11 in \citealt{Cenko2011}). The values for $\theta_0$ for the case of no jet break indicate a spherical outflow, opposite to the collimated outflow usually expected and assumed in the standard afterglow theory. Finally, the energy lower limit is $\sim 3 \times 10^{52}$ erg, implying an efficiency of $\eta < 70\%$.

\subsection{Origin of the light-curve break}

From the available data, it is difficult to make a clear case for a preferred model for this GRB afterglow. Each of the studied models has specific problems that are difficult to explain with a simple afterglow model and would probably be better understood with a more complex and detailed model of the afterglow emission, especially at early times (e.g. \citealt{2003ApJ...584..390W,2007ApJ...665..569M}). However, we are able to rule out some of the possible models. For instance, the jet-break model where the spectral regime is $\nu_{\rm{sa}} < \nu_{\rm{m}}$ is ruled out because the spectral evolution will never cross that regime in the slow cooling phase (see Sect. \ref{jetbreak}). In a similar way, we can rule out the energy-injection model with a wind density profile $k=2$, and the energy-injection model with a jet break for $k=0$ because $\bar{\epsilon}_{\rm{e}}$ has to be larger than one, which is physically not meaningful. The energy-injection model without jet break for $k=1.1$ and ISM density profiles cannot be ruled out. However, in this model the extremely high density requirements are far from theoretically expected values and previous measurements. Moreover, the resulting spherical outflow geometry, implied by the derived value for the half-opening angle, would require a very energetic explosion. \\

\begin{figure}[h!]
\includegraphics[width=0.5\textwidth]{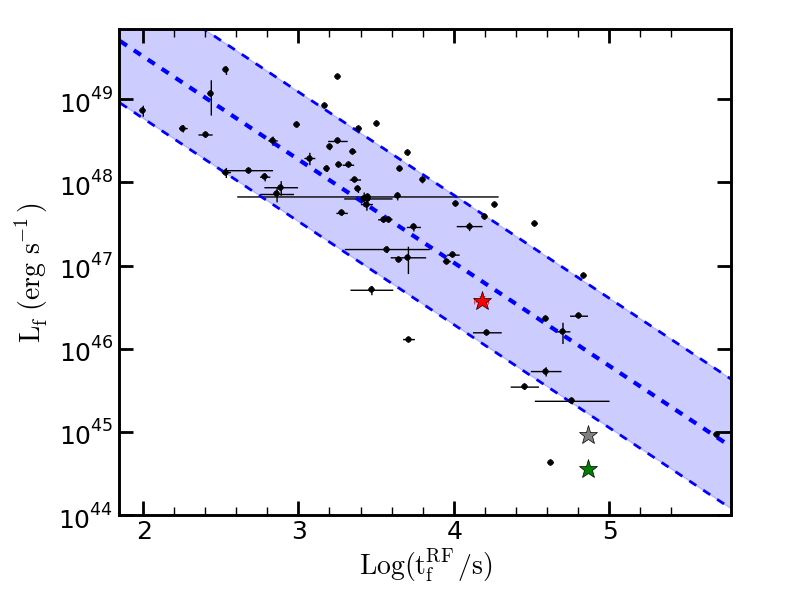}
\caption{$0.3-30$ keV luminosity at the end of the energy-injection phase \citep{2008MNRAS.391L..79D}. The black dots are taken from \citet{2013MNRAS.428..729M}. The stars correspond to the GRB 121024A afterglow: the red star to when the end of the plateau phase lies at 49.8 ks, the green star corresponds to the ongoing energy-injection phase before and after the break in the LC at t = 240 ks, and the grey star corresponds to the luminosity corrected for the jet-break effect. The dashed line in the middle corresponds to the best fit, and the shaded region is the $1\sigma$ error of the fit.}
\label{margutti7}
\end{figure}

For more than $40\%$ of the X-ray afterglows an initial plateau lasting for about $10^3-10^4$ s is observed \citep{2007MNRAS.375L..46L,2013MNRAS.428..729M} and has been associated with a continuous energy injection during the afterglow evolution \citep{2006nousek,2014MNRAS.442.3495V}. We compared the X-ray luminosity ($0.3-30$ keV) ($L_f$) and the break time in the rest frame ($t^{RF}_{f}$) to the relation observed in other GRBs \citep{2008MNRAS.391L..79D}. Figure \ref{margutti7} shows the relation between $L_f$ and $t^{RF}_{f}$observed in a sample of 62 long GRBs studied by \citet{2013MNRAS.428..729M}. We included GRB 121024A for both an energy-injection phase that ends at the time of the break in the light curve (red star) and for an ongoing injection phase until the end of the observations at $t = 240$ ks, both taking the end-time luminosity directly (green star) and correcting for the change in the slope introduced at the break (grey star). Energy injection follows the correlation very well, supporting the scenario with energy injection up to the break in the light curve. Continued energy injection is disfavoured in view of the correlation. We note that assuming energy injection to extend beyond the final data point at 240 ks will only shift the grey and green stars even farther away from the correlation.\\

We consider the jet-break model where $\nu_{\rm{m}} < \nu_{\rm{sa}}$ to be the preferred scenario. In this model, all the micro-physical and dynamical parameters are within the range of previous measurements and within the expected values from the standard afterglow model. The low values for the energy are just lower limits and therefore are not a strong argument against this model. The main problem is related to the hard electron spectrum that requires additional assumptions on the acceleration process of the electrons in the shock region. However, this is certainly not the first GRB for which such a shallow electron spectrum was derived, and viable ways to handle this scenario have been put forward, two of which we investigated and found to give reasonable and physically meaningful results. The derived hard electron spectrum need not be a reason to reject a model, and more likely reflects our poor understanding of acceleration processes under extreme conditions. Finally, the linear polarisation observations reported by \citet{2014Natur.509..201W} would agree with a jet-break model where the linear polarisation would be a direct result from the jet break. However, there are still no studies reported in the literature analysing whether it would be possible to obtain this type of polarisation from an energy-injection model. 

\section{Summary and conclusions}
\label{sum}

We analysed the afterglow of GRB 121024A and showed that the multi-wavelength data enable the afterglow spectra and temporal parameters to be measured to a high degree of accuracy, which sets strong constraints on the micro-physics in the shock region and on the dynamics of the jet. The combined GROND and XRT data allowed us to determine the spectral slope $\beta$ in this energy regime with high accuracy, and therefore we were able to measure the electron index $p$. We modelled our complete set of observations using two different physical interpretations: a jet-break model and an energy-injection model. The energy-injection model requires $\eta < 77\%$, $71\%,$ and $69\%$ for k = 2, 1.1, and 0, respectively, and does not contradict Fermi acceleration predictions for the electron index $p$. However, it does face some problems with the derived microphysical parameters in the case of a wind density profile, and the density values are extremely high in all three of the density profiles studied. \\

The jet-break model requires a hard electron spectrum. Assuming $\gamma_m \propto \gamma$, the derived microphysical and dynamical parameters are all consistent with previous measurements and with expected values from theoretical analysis. There is a problem with the efficiency requirements, which for $\nu_{\rm{sa}} < \nu_{\rm{m}}$ can be as high as $\eta\sim98\%,$ and for $\nu_{\rm{m}} < \nu_{\rm{sa}}$ the efficiency has an upper limit of $\eta < 71\%$.\\

The results presented here on GRB 121024A show that broadband afterglow data from the X-ray to the radio allow for a detailed analysis of the characteristic properties of the GRB afterglow synchrotron emission spectrum. As studies of other GRBs have also shown, such datasets are invaluable for determining the range of microphysical and dynamical parameters within GRB shock-fronts with better statistics and avoiding adding additional assumptions to the analysis. Through our extensive data coverage of GRB 121024A, we have been able to constrain the position of all synchrotron breaks, which in turn has allowed us to measure, or set constraints on, all the micro-physical and dynamical parameters of a GRB afterglow. This information is crucial to further  study the GRB afterglow emission process, and we are currently working on a larger sample of GRBs with sensitive and broadband afterglow data (Varela et al. in preparation). Future continual coverage of the GRB afterglows with sensitive telescopes over a wide wavelength range and at multiple epochs will enable us to place strong constraints on the micro-physical parameters for a larger sample of GRBs, and allow us for instance to investigate the evolution of these parameters.

\begin{acknowledgements}
We express special thanks to Reg\'{i}s Lachaume for helping with the GROND observations. We are grateful to Rafaela Margutti for the data of Fig. \ref{margutti7}. K.V. acknowledges support by DFG grant SA 2001/2-1. H.v.E.  acknowledges support by the Alexander von Humboldt Foundation and the Carl Friedrich von Siemens Foundation. J.G. acknowledges support by the DFG cluster of excellence "Origin and Structure of the Universe" (www.universe-cluster.de). R.F. acknowledges support by Czech MEYS Grant 7AMB14DE001. P.S. acknowledges support through the Sofja Kovalevskaja Award from the Alexander von Humboldt Foundation of Germany. J.F.G., M.T. and P.W. acknowledge support through the Sofja Kovalevskaja Award to P. Schady from the Alexander von Humboldt Foundation of Germany. S.K. and A.N.G. acknowledge support by DFG grant Kl 766/16-1. A.N.G. and D.A.K. are grateful for travel funding support through MPE. S.S. acknowledges support by the Th\"{u}ringer Ministerium f\"{u}r Bildung, Wissenschaft und Kultur under FKZ 12010-514. A.Rossi acknowledges support by the Th\"{u}ringer Landessternwarte Tautenburg. C.D. acknowledges support through EXTraS, funded from the European Union's Seventh Framework Programme for research, technological development and demonstration under grant agreement no 607452. APEX is operated by the Max-Planck-Institut f\"{u}r Radioastronomie, the European Southern Observatory, and the Onsala Space Observatory. Part of the funding for GROND (both hardware as well as personnel) was generously granted from the Leibniz-Prize to  G. Hasinger (DFG grant HA 1850/28-1). This work made use of data supplied by the UK \textit{Swift} Science Data Centre at the University of Leicester.
\end{acknowledgements}

\appendix
\section{Alternative analysis to the jet-break model}

One of the interpretations for the break in the light curve is a jet break. As mentioned in Sect. \ref{jetbreak}, a hard electron spectrum is required in this scenario. For completeness, we present
here a second approach to analyse a hard electron spectrum \citep{2001BASIBBhattacharya,DC2001,Gao2013141}. Here, instead of assuming $\gamma_{\rm{m}} \propto \gamma$, the effect of an upper cut-off $\gamma_{\rm{M}}=[3\rm{e}/\phi \sigma_{\rm{T}} B]^{1/2}$ in the energy range of the accelerated particle population can be included in the minimal Lorentz factor such that $\gamma_{\rm{m}}\propto(\gamma \gamma_{\rm{M}}^{(p-2)})^{(1/p-1)}$ \citep{DC2001}. This upper cut-off follows from equating acceleration and synchrotron cooling timescales. The advantages of this approach are that the extra cut-off is modelled explicitly and that $\epsilon_e$ can still be interpreted as the fraction of energy in the accelerated electrons. However, it implies that the behaviour of the electron population at low energies is dictated by the few electrons that were accelerated more efficiently, which is not supported by any clear physical mechanism.\\

We calculated all the parameters for both regimes, when $\nu_{\rm{sa}} < \nu_{\rm{m}}$ and when $\nu_{\rm{m}} < \nu_{\rm{sa}}$, and report them in Table \ref{paramstableDC}. The transition from the fast to slow cooling regime occurs at $t_{\nu_c = \nu_m} \sim 3.6\times10^4$ s for $\nu_{\rm{sa}} < \nu_{\rm{m}}$ and $t_{\nu_c = \nu_m} \sim 6.6\times10^2$ s for $\nu_{\rm{sa}} > \nu_{\rm{m}}$, consistent with the slow cooling assumption. The transition from optically thin to optically thick occurs at $t_{\nu_{sa} = \nu_m} \sim  9.8 \times10^4$ s for $\nu_{\rm{sa}} < \nu_{\rm{m}}$ and at $t_{\nu_{sa} = \nu_m} \sim  1.2 \times10^4$ s for $\nu_{\rm{sa}} > \nu_{\rm{m}}$.  In this case, the sub-regime where $\nu_{\rm{sa}} < \nu_{\rm{m}}$ is ruled out because here the time that $\nu_{\rm{m}}$ would cross $\nu_{\rm{sa}}$ is $\sim12$ ks, which is before the epoch of the SED ($109$ ks) used in the analysis. In the second sub-regime, where $\nu_{\rm{m}} < \nu_{\rm{sa}}$, $\theta_0 $ is consistent with a collimated outflow and $A_{*}$ is in the range of expected values for a wind environment and corresponds to a mass-loss rate of $> 2.6\times10^{-6} \, \dot{\rm{M}}_o \, \rm{yr}^{-1}$ for a wind velocity $v = 1000$ km s$^{-1}$ \citep{1999ApJ...520L..29C,2000ApJ...536..195C}, consistent with a Wolf-Rayet star as a possible progenitor.  The efficiency $\eta < 98\%$, even though it is just an upper limit, is extremely high. Our results agree with the results in Sect. \ref{jetbreak} and support the jet-break scenario when $\nu_m < \nu_{sa}$ as the preferred scenario.

\begin{table}[!ht]
\centering\setlength\tabcolsep{5.0pt}
\renewcommand*{\arraystretch}{1.2}
\caption{$\gamma_{\rm{m}}$,$\bar{\epsilon}_{\rm{e}}$, $\epsilon_{\rm{B}}$, $E_{\rm{iso}}$, $n$ and $\theta_0$ for a jet-break model. $p=1.73\pm0.03$. $\nu_{\rm{c}} < \nu_{\rm{K_s}}$}
\label{paramstableDC}
\begin{tabular}{c|c|c|c}
\toprule
\toprule
\multicolumn{1}{c}{} & $\bar{\epsilon}_{\rm{e}}$& $\epsilon_{\rm{B}}$ & $A_{*}$ \\
\midrule
$\nu_{\rm{sa}} < \nu_{\rm{m}}$   & $0.80^{+0.20}_{-0.62}$ & $7.46^{+1.33}_{-6.37}\cdot10^{-3}$ & $2.07^{+3.46}_{-1.38}$  \\
$\nu_{\rm{m}} < \nu_{\rm{sa}}$    & < $0.11$ & < $0.18$ & > $0.26$ \\
\midrule
\multicolumn{1}{c}{} & $\theta_0$ [rad]  & $E_{\rm{iso},52}$[erg] & $\,$ $\eta$ $\,$\\
\midrule
$\nu_{\rm{sa}} < \nu_{\rm{m}}$   &  $3.75^{+18.33}_{-0.62}\cdot10^{-2}$ & $1.25^{+0.75}_{-0.75}\cdot10^{-2}$ & $99^{+1}_{-8}\%$ \\
$\nu_{\rm{m}} < \nu_{\rm{sa}}$   &  > $1.12\cdot10^{-2}$ & > $0.19$ &  $<98\%$\\

\bottomrule
\end{tabular}
\end{table}

\bibliographystyle{aa}
\bibliography{biblio}


\end{document}